\documentclass[twocolumn,aip,jcp,amsmath,amssymb,10pt]{revtex4-1}
\usepackage{graphicx,amsmath}
\begin{document}
\title{High-harmonic transient grating spectroscopy of NO$_2$ electronic relaxation}
\author{H. Ruf}
\affiliation{Universit\'e de Bordeaux - CNRS - CEA, CELIA , UMR5107, F-33400 Talence, France}
\author{C. Handschin}   
\affiliation{Universit\'e de Bordeaux - CNRS - CEA, CELIA , UMR5107, F-33400 Talence, France}
\author{A. Ferr\'e}
\affiliation{Universit\'e de Bordeaux - CNRS - CEA, CELIA , UMR5107, F-33400 Talence, France}
\author{N. Thir\'e}
\affiliation{Laboratoire Collisions Agr\'egats R\'eactivit\'e-IRSAMC, UPS, Universit\'e de Toulouse and CNRS, UMR 5589, F-31062 Toulouse, France}
\author{ J. B. Bertrand} 
\affiliation{Joint Attosecond Science Laboratory, National Research Council of Canada and University of Ottawa 100 Sussex Drive, Ottawa, ON Canada K1A 0R6}
\author{L. Bonnet}
\affiliation{ Universit\'e de Bordeaux - CNRS, ISM , UMR5255, F-33400 Talence, France}
\author{R. Cireasa}
\affiliation{Laboratoire Collisions Agr\'egats R\'eactivit\'e-IRSAMC, UPS, Universit\'e de Toulouse and CNRS, UMR 5589, F-31062 Toulouse, France}
 \author{E. Constant}
\affiliation{Universit\'e de Bordeaux - CNRS - CEA, CELIA , UMR5107, F-33400 Talence, France}
\author{P. B. Corkum}
\affiliation{Joint Attosecond Science Laboratory, National Research Council of Canada and University of Ottawa 100 Sussex Drive, Ottawa, ON Canada K1A 0R6}
\author{D. Descamps}
\affiliation{Universit\'e de Bordeaux - CNRS - CEA, CELIA , UMR5107, F-33400 Talence, France}
\author{B. Fabre}
\affiliation{Universit\'e de Bordeaux - CNRS - CEA, CELIA , UMR5107, F-33400 Talence, France}
\author{P. Larregaray}
\affiliation{ Universit\'e de Bordeaux - CNRS, ISM , UMR5255, F-33400 Talence, France}
\author{E. M\'evel}
\affiliation{Universit\'e de Bordeaux - CNRS - CEA, CELIA , UMR5107, F-33400 Talence, France}
\author{S. Petit}
\affiliation{Universit\'e de Bordeaux - CNRS - CEA, CELIA , UMR5107, F-33400 Talence, France}
\author{B. Pons}
\affiliation{Universit\'e de Bordeaux - CNRS - CEA, CELIA , UMR5107, F-33400 Talence, France}
\author{D. Staedter}
\affiliation{Laboratoire Collisions Agr\'egats R\'eactivit\'e-IRSAMC, UPS, Universit\'e de Toulouse and CNRS, UMR 5589, F-31062 Toulouse, France}
\author{H. J. W\"orner}
\affiliation{Joint Attosecond Science Laboratory, National Research Council of Canada and University of Ottawa 100 Sussex Drive, Ottawa, ON Canada K1A 0R6}
\author{D. M. Villeneuve}
\affiliation{Joint Attosecond Science Laboratory, National Research Council of Canada and University of Ottawa 100 Sussex Drive, Ottawa, ON Canada K1A 0R6}
\author{Y. Mairesse}
\affiliation{Universit\'e de Bordeaux - CNRS - CEA, CELIA , UMR5107, F-33400 Talence, France}
\author{P. Halvick }
\affiliation{ Universit\'e de Bordeaux - CNRS, ISM , UMR5255, F-33400 Talence, France}
\author{V. Blanchet}
\affiliation{Laboratoire Collisions Agr\'egats R\'eactivit\'e-IRSAMC, UPS, Universit\'e de Toulouse and CNRS, UMR 5589, F-31062 Toulouse, France}

\begin{abstract}
We study theoretically and experimentally the electronic relaxation of NO$_2$ molecules excited by absorption of one $\sim$ 400 nm pump photon. Semi-classical simulations based on trajectory surface hopping calculations are performed. They predict fast oscillations of the electronic character around the intersection of the ground and first excited diabatic states. An experiment based on high-order harmonic transient grating spectroscopy reveals dynamics occuring on the same timescale. A systematic study of the detected transient is conducted to investigate the possible influence of the pump intensity, pump wavelength, and rotational temperature of the molecules. The quantitative agreement between measured and predicted dynamics shows that, in NO$_2$, high harmonic transient grating spectroscopy encodes vibrational dynamics underlying the electronic relaxation.
\end{abstract}
\maketitle
\normalsize\section{Introduction}
Resolving in time the energy relaxation in molecules both at the vibrational and electronic levels is theoretically and experimentally challenging. There are a variety of experimental techniques for probing molecular dynamics, whose utility can vary from one polyatomic molecule to another one \cite{lhermite, Minitti,davies,davies2}. Among the most commonly used pump-probe signals, we can highlight photoelectron energy \cite{stolow2003} and angular distributions (PAD) \cite{suzuki}, electron-ion coincidence in the perturbative regime \cite{jiang, vredenborg} that can provide molecular frame-PAD \cite{davies, Hockett}, or in the non-perturbative regime like Coulomb explosion \cite{Hishikawa} or above threshold ionization \cite{Spanner:12}. All these detections are based on collecting charged species. An alternative experimental technique developped in the last decade is high harmonic spectroscopy, based on the high order harmonic generation process. This method has shown a relevant sensitivity to nuclear dynamics \cite{Itatani04,wagner,li,worner2} and an inherent connection to the symmetries of molecular orbitals \cite{Soifer,wong,wong2, Frumker:12}. 

The high harmonic generation (HHG) process, which occurs when a molecule is submitted to a strong probe laser field, can be described in a first approximation by a three step model \cite{Lewenstein}. A valence electron first tunnels out of the molecule under the influence of the strong field.
This electron is then driven away by the electric field and can finally recombine onto its ionic core to emit extreme ultraviolet radiation. This process occurs every half period of the probe laser, so that the resulting radiation is constituted of odd high-order harmonics. The amplitude, polarization and phase of the XUV-VUV emission encode information about the electronic configuration of the target molecules. 

Since this pump-probe signal is purely based on optical detection, its contrast can be drastically reduced due to the harmonic emission from unexcited molecules. This can be improved by using for instance polarization resolved spectroscopy \cite{Mairesse08polar} or transient grating spectroscopy \cite{mairesse}. In the latter case, a spatial grating of molecular excitation is created by optical interferences between two identical non-colinear pump pulses. The sinusoidal spatial modulation of the pump intensity across the molecular beam results in a partial diffraction of the high harmonic emission. 

\begin{figure}
\begin{center}
\includegraphics[width=0.4\textwidth]{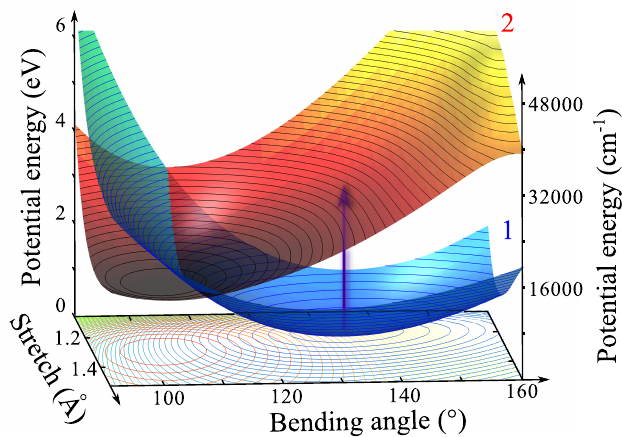}
\end{center}
\caption{Potential energy surfaces of the diabatic fundamental state 1$^2$A' and first excited state 2$^2$A' of NO$_2$ as a function of the bending angle and NO strecth distance R$_1$. These diabatic states are referred to as 1 and 2 in the text, respectively. The second NO distance $R_2$ is fixed to 1.19 \AA.}
\label{FigSurfaces}
\end{figure}

This approach has been used in the present paper to revisit the relaxation dynamics in NO$_2$ molecules electronically excited in the first optically-allowed electronic state $^2$B$_2$ close to the first dissociation limit ($\sim$ 3 eV, Fig. \ref{FigSurfaces}). The dipole of this transition belongs to the B$_2$ irreducible representation of the C$_{2v}$ point group, namely the pump pulses preferentially excite NO$_2$ molecules with the C$_{2v}$  axis perpendicular to the pump polarization. By reducing the point group symmetry to C$_s$, the 1$^2$A$_1$ and 1$^2$B$_2$ states become the 1$^2$A' and 2$^2$A' states respectively that are vibronically coupled. We call these diabatic states 1 and 2, respectively, as labelled in Fig. 1. This strong coupling results in a high density of states (0.754 levels/cm$^{-1}$) \cite{ionov} and leads to the loss of both the electronic and vibration/rotation signatures. This has made of NO$_2$ a kind of hellish grail of molecular physicists. Time-resolved photoionization has revealed unassigned regular oscillations with a 500 fs period for a pump centered at 400 nm \cite{Form,Cireasa,Hamard,vredenborg}. 

On the contrary, a recent high harmonic spectroscopy experiment using $\sim 400$ nm pump pulses has shown a few oscillations of $\sim$ 100 fs characteristic time that have been assigned to the $^2$B$_2$-$^2$A$_1$ vibronic coupling \cite{worner2011}, in qualitative agreement with theoretical predictions \cite{arasaki07,arasaki10}. In the present paper, we reproduce this experiment and find subtle but significant differences in the femtosecond transient with respect to previous results \cite{worner2011}. We investigate the influence of the pump intensity and wavelength, and of the rotational temperature of the molecular beam. Simulations based on Trajectory Surface Hopping (TSH) reveal that the diffracted signal from the excitation grating directly reflects the vibrational dynamics of the first excited diabatic state around the conical intersection. 
The remainder of the paper is organized as follows. Calculations of the molecular dynamics are presented in Section II. The experimental results of high harmonic transient grating spectroscopy are given in Section III. Theory and experiments are combined to deliver an interpretation to the observed dynamics in Section IV. 

\section{Simulations}
\subsection{Background}

NO$_{2}$ has received much attention because it is a stable and small molecule for which the effect of a conical intersection can be observed and calculated. Most of the theoretical studies were focused on the analysis of the experimental spectra and were performed in a time-independent approach. More recently, time-dependent simulations of the intramolecular nonadiabatic dynamics were also performed \cite{Santoro:99,Santoro:00,Sanrey:06,Arasaki:07}. These studies were based on a diabatic representation of the electronic Hamiltonian obtained from \textit{ab initio} calculations \cite{Santoro:99,Santoro:00,Arasaki:07} or from the experimental spectrum \cite{Sanrey:06}. By propagating a quantum wavepacket or a swarm of surface hopping trajectories, the time dependent populations of diabatic or adiabatic states were calculated and regular recurrences were observed in some of these populations. For a range of total energy going from $\sim$ 13000 cm$^{-1}$ up to $\sim$ 28000 cm$^{-1}$, the period of these 
recurrences is in the range 50 - 100 fs with a propensity to increase when total energy increases. A good agreement was observed between quantum time-dependent (QTD) and TSH results \cite{Santoro:00,Sanrey:06}, indicating that the semiclassical approach is relevant for this system.

Since experiments are carried out around 3 eV, the comparison with theory requires accurate PES including the O+NO dissociation channel. Among the three existing models of PES\cite{Leonardi:96,Arasaki:07,Kurkal:03} satisfying these requirements, the most reliable one was built by Kurkal, Fleurat-Lessard and Schinke\cite{Kurkal:03} (KFS) from \textit{ab initio} data calculated with multireference configuration interaction and a quadruple zeta basis set. This last work provides both the adiabatic and diabatic representations of the ground and the first excited PES. The bound levels of the ground electronic state were calculated up to $\sim$ 7050 cm$^{-1}$ and a good qualitative agreement with the experimental spectrum was obtained. 

However the KFS diabatic representation is incomplete because neither the diabatic electronic coupling nor the nonadiabatic derivative coupling have been calculated. Nevertheless the square of the electronic coupling $V_{12}$ can be extracted from the available diabatic ($E_{1}$ and $E_{2}$) and adiabatic ($E_{X}$ and $E_{A}$) energies by
\begin{equation} \label{eq:v12carre}
V_{12}^{2} = \frac{1}{4}[(E_{A}-E_{X})^{2}-(E_{2}-E_{1})^{2}]
\end{equation}
Let $Q_{-}$ = $R_{1}-R_{2}$ be the antisymmetric stretch coordinate, $R_{1}$ and $R_{2}$  being the N-O distances. When $Q_{-}$ = 0, the molecule has the $C_{2v}$ symmetry, the electronic coupling should be equal to zero and the adiabatic energies should be equal to the diabatic energies. If $Q_{-} \neq$ 0, the symmetry is reduced to $C_{S}$, the electronic coupling can be different from zero and the following inequality must be fullfiled:
\begin{equation} \label{eq:inegal}
|E_{A}-E_{X}| \geq |E_{2}-E_{1}|
\end{equation}
Because the adiabatic and diabatic PES have been separately fitted, the inequality (\ref{eq:inegal}) is not always satisfied due to the small errors introduced by the fitting procedures. Since this occurs in regions where $V_{12}$ is close to zero, we safely reduce it to exactly zero. The sign of $V_{12}$ cannot be known from equation (\ref{eq:v12carre}), but at the first order, $V_{12}$ is proportional to $Q_{-}$ \cite{Carrington:74,Leitner:96}. For this reason which is a particular case of a more general symmetry reason \cite{Hirsch:91}, $V_{12}$ takes the sign of $Q_{-}$.

Finally, let us remind two important energy quantities of the PES. The O+NO dissociation energy of the ground electronic state is $D_{0}$ = 3.00 eV  while the experimental value is $D_{0}$ = 3.11 eV \cite{jost96}. The difference of potential energy from the ground to the excited electronic state, both at equilibrium configuration, is $T_{e}$ = 1.31 eV, while the experimental value is 1.21 eV \cite{Delon:01}.

\subsection{Trajectory surface hopping calculations}

Although the nonadiabatic dynamics of a triatomic molecule can be simulated nowadays with a QTD method without difficulties, we have preferred the semiclassical approach. The obvious benefit of using semiclassical methods is the possibility to study large polyatomic systems for which quantum dynamics calculations are out of reach. Validating the semiclassical approach for the small system NO$_{2}$ will pave the way to the study of larger systems. While TSH simulations of the dynamics of NO$_{2}$ have been already successfully compared to QTD simulations \cite{Santoro:00,Sanrey:06}, the present work provides the first comparison of the former with experimental data. 
The TSH method used here is based on the Tully's fewest switches algorithm \cite{Tully:90} and some implementation details can be found in a previous work \cite{Halvick:02}.

The starting point in the phase space and the starting time of the trajectories are obtained by applying the classical treatment proposed by Meier and Engel\cite{Meier:02}. The initial phase space distribution is obtained by a two-step procedure. First, the initial distribution for the vibrational ground state of the electronic ground state, represented by the Wigner function $\rho_{W}^{(X)}$, is filtered by the resonant one-photon transition
\begin{equation}
\rho(\mathbf{q},\mathbf{p}) = \rho_{W}^{(X)}(\mathbf{q},\mathbf{p})      e^{-(D_{12}-\hbar\omega)^{2}/2\alpha\hbar^{2}}
\end{equation}
where $D_{12}$ is the potential energy difference of the transition, $\omega$ the central frequency of the pulse, and $\alpha$  the parameter of the pulse envelope defined by $e^{-\alpha t^{2}}$. Second, the random starting time of each trajectory is drawn from the distribution function $e^{-2\alpha t^{2}}$. The quantities $\omega$ and $\alpha$ have been defined by considering a pump pulse centred at 400 nm and with a 30 fs FWHM gaussian envelope.

\begin{figure}
\begin{center}
\includegraphics[width=0.5\textwidth]{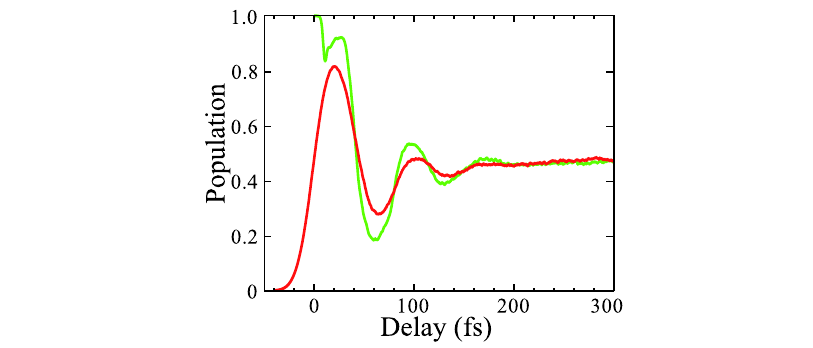}
\end{center}
\caption{ Temporal evolution of the $P_{2}$ diabatic population  calculated with a batch of trajectories starting at the same time (green) or starting at gaussian random time (red), in the case of a 30 fs pump pulse centred at 400 nm.}
\label{Fig_pop2}
\end{figure}

Figure \ref{Fig_pop2} shows two different evolutions of the population $P_{2}$ of the excited diabatic state: the first one is calculated by applying only the first step of the initial procedure, and the second one is calculated with the complete initial procedure. As expected, using random starting time tends to smooth out the details of the dynamics. 
These calculations have been done by propagating a batch of $1\times10^{4}$ trajectories. Smaller batches of $2\times10^{3}$ and $5\times10^{3}$ result in no observable difference in the $P_{2}$ population behavior, except that some numerical noise increases progressively as the number of trajectories decreases.
The influence of the pump wavelength was also investigated by computing $P_{2}$ for a laser pulse centred at 395 nm and at 392 nm. No significant difference could be observed with respect to the result at 400 nm shown in Fig \ref{Fig_pop2}. All three curves were almost indistinguishable until $t = 100$ fs.

\begin{figure}
\begin{center}
\includegraphics[width=0.4\textwidth]{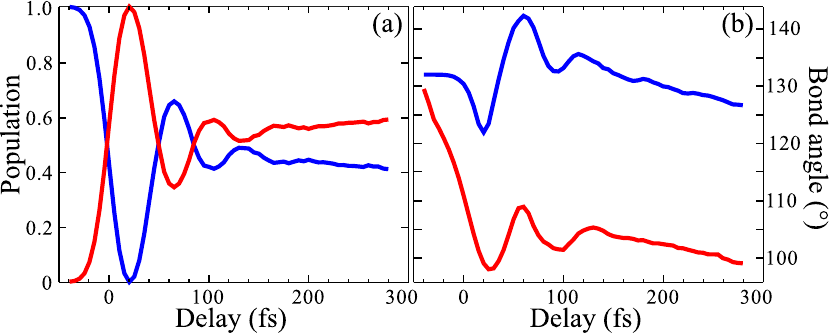}
\end{center}
\caption{Temporal evolution of the ground (blue) and first excited (red) diabatic state populations (a) and average bond angle (b) from TSH calculations, assuming a 100$\%$ excitation probability.}
\label{FigPopAngles}
\end{figure}

\begin{figure}
\begin{center}
\includegraphics[width=0.6\textwidth]{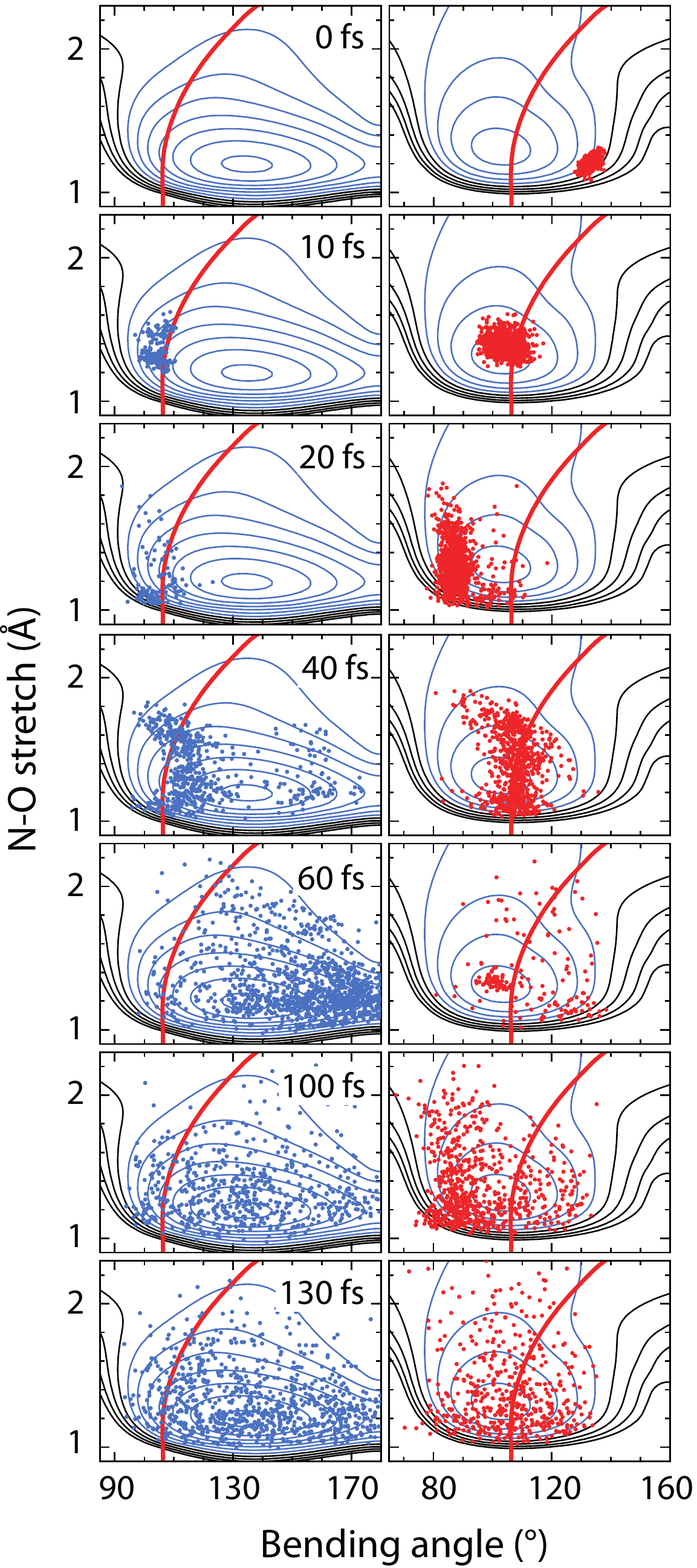}
\end{center}
\caption{Snapshots of the trajectories swarm and contour lines of the diabatic PES. The blue and red dots represent the coordinates of the trajectories on the ground and excited diabatic PES respectively. The contour lines are separated by 0.5 eV and the blue contour lines are below or equal to the O+NO dissociation limit. The crossing seam is represented by the thick red line. The PES are represented as a function of the bending angle and the NO distance $R_{1}$, with the second NO distance $R_{2}$ fixed to 1.19 \AA. The snapshots are obtained by representing only two coordinates (bending angle and $R_{1}$) and ignoring the third one ($R_{2}$).}
\label{Fig_snapshot}
\end{figure}

Let us now examine the intramolecular nonadiabatic dynamics revealed by these calculations. Figure \ref{FigPopAngles} shows the evolution of the populations $P_{1}$ and $P_{2}$ on the ground and excited diabatic PES respectively, along with the evolution of the average bond angle on each PES. Additionally, Figure \ref{Fig_snapshot} displays some snapshots  of the trajectories swarm on both diabatic surfaces at selected time steps. Because the details of the dynamics are more easily observed in the case where random initial delay is not applied, the snapshots where obtained from a batch of trajectories starting at the same time on the excited PES. Only $2\times10^{3}$ trajectories are included in order to avoid overloading the figure. At the beginning, $P_{1}(t=0)$ = 0 and $P_{2}(t=0)$ = 1. Then the swarm goes downhill and encounters the crossing seam at $t \sim10$ fs. Because the swarm is still quite close to a  $C_{2v}$ geometry, the electronic coupling is inefficient, with few trajectories jumping on the ground PES. The swarm reaches the turning point at small bond angles at $t \sim20$ fs, spreads along Q$_-$ before returning to the crossing seam. Due to this Q$_-$ dispersion, the electronic coupling is effective at this second crossing of the seam. Therefore, a large number of trajectories jumps on the ground state. The main effect is the drastic decrease of $P_{2}$ observed between $t$ = 30 fs and $t$ = 60 fs on Figure 3 (a). The variation of $P_{2}$ is now controlled by the bending motion of the swarm on the ground PES. This swarm moves to the O-N-O linear configuration ($t$ = 60 fs) and then goes back towards small angles, crosses again the seam and thus repopulates the excited PES: this is the first recurrence of $P_{2}$ at $t \sim100$ fs. Another period of bending occurs, indicated by a second recurrence which is hardly visible around $ \sim170$ fs. Then both swarms are completely scattered in the whole phase space and no more change of the populations can be seen.

In the adiabatic representation, the dynamics is somewhat simpler. The trajectories start on the excited adiabatic state. Around t=10 fs, most of the trajectories jump on the ground adiabatic state and then remain in that state, in which a large amplitude bending motion occurs subsequently. Thus, the same dynamics is seen as a variation of the population in the diabatic representation or as a variation of the average bending angle in the adiabatic representation.  

The time dependencies of the diabatic populations (Fig. \ref{FigPopAngles}(a)) as well as the snapshots of the trajectories (Fig \ref{Fig_snapshot}) will be used in section IV to analyze the results of the experiment presented in the following section.

\section{High harmonic transient grating spectroscopy}

Transient grating spectroscopy is widely used in non-linear optical spectroscopy to perform background-free measurements. In this configuration, the molecular medium is excited by two synchronized non-colinear beams which optically interfere, forming a grating of molecular excitation.  This technique can be directly transposed to the extremely non-linear regime of high-order harmonic spectroscopy \cite{mairesse}. The probe beam generates high harmonics in this structured medium. The harmonics are diffracted by the grating of excitation, leading to the appearance of one or several diffraction peaks around the main harmonic beam. In the present experiment, the fringes of the excitation grating being horizontal the diffraction peaks appear above and below the main harmonic beam. This diffraction pattern can be used to perform high contrast pump-probe measurements, since there is no signal in the direction of the diffracted peaks in the absence of molecular excitation. In addition, since the diffraction pattern is formed by the interference of emission from excited and unexcited sources, it encodes the phase difference between harmonic emission from excited and ground state molecules \cite{mairesse2,worner2}. 

High harmonic transient grating spectroscopy is particularly valuable in a context where the pump can induce dynamics both through one-photon and multi-photon absorption. The spatial modulation of the pump beam intensity is sinusoidal. For a pure one-photon process, the excitation grating is also sinusoidal and only first order diffraction can be observed in the far field (Fig. \ref{FigTGS}(a)). If the excitation results from a two-photon process, then the grating is determined by the square of the sinusoidal modulation of the pump beam, which contains the second harmonic of the exciting beam spatial modulation frequency. Consequently, second order diffraction peaks appear (Fig. \ref{FigTGS}(b)). Note that second (and higher) order diffraction peaks can also appear in the case of a saturated one-photon transition: the saturation of the molecular excitation induces an anharmonicity of the grating which is reflected in the far field diffraction pattern (Fig. \ref{FigTGS}(b)). 

\subsection{Experimental setup}
\begin{figure}
\begin{center}
\includegraphics*[width=0.45\textwidth]{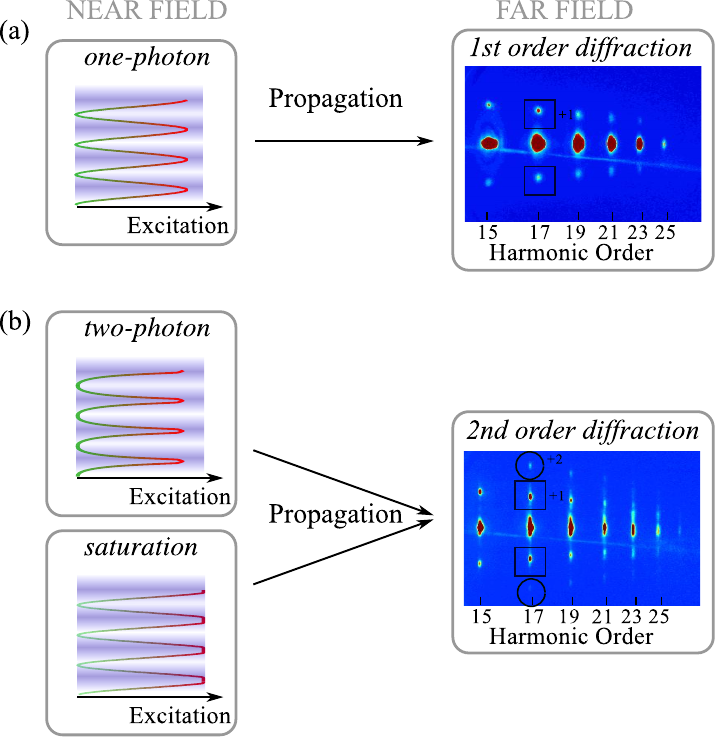}
\end{center}
\caption{Principle of high harmonic transient grating spectroscopy. In the case of one-photon excitation (a), the sinusoidal modulation of the pump intensity in the near field produces a sinusoidal modulation of the harmonic emission, which results in the appearance of first order diffracted light up and down the harmonics in the far field. The far field profile was experimentally recorded by using two pump pulses containing 20 $\mu$J/p, at a pump-probe delay of 100 fs and with a probe pulse polarized orthogonally to the pump pulses. At higher pump intensity (35 $\mu$J/p, (b)), second order diffraction peaks appear in the far field. They can result from two-photon transitions or from the saturation of the one-photon absorption, the two effects leading to anharmonicities of the near-field harmonic spatial modulation.}
\label{FigTGS}
\end{figure}

The experiment was performed using the Aurore Ti:Sa source at CELIA, delivering 7 mJ, 27 fs, 800 nm pulses at 1 kHz. The laser beam is split into a pump (20$\%$ in energy) and probe (80$\%$ in energy) beams. The pump beam is frequency doubled using a 200 $\mu$m SH-BBO crystal, which provides wavelength tunability around 400 nm without reducing the temporal resolution ($\sim$ 35 fs corresponding to a 7 nm bandwidth). The pump is further split in two parts by a 50-50 beamsplitter. The 800 nm light is filtered out from the pump beams using two dichroic dielectric mirrors (in each arm) with high reflectivity around 400 nm and high transmission around 800 nm. An adjustable aperture on the pump beams is used to reduce their energy and to ensure that the excitation of the molecular medium will be uniform over the area probed by high-harmonic generation. The pump energy used in the experiment typically varied between 10 and 35 $\mu$J/p in each arm. The two pump beams are parallel and vertically shifted with respect to each other by 16 mm. 
The 800 nm probe beam is centered between the two pumps. The three beams are focused onto the gas target using 750 mm radius spherical silver mirror placed under vacuum in order to minimize self phase modulation in the entrance window of the vacuum chamber. The pump beams produce an interference pattern at focus with a fringe spacing of 9.4 $\mu$m. The probe pulse waist is $\sim$35 $\mu$m. The linear polarization of the probe pulse can be continuously rotated relative to the S-polarized pump pulses. The typical probe intensity determined from the harmonic cut-off (see Fig \ref{FigTGS}) is $\sim$1.2$\times$10$^{14}$ W/cm$^2$. The high-harmonic emission is produced a few 100 $\mu$m downstream from a 100 $\mu$m nozzle gas jet (General Valve) operated at 100 Hz with a backing pressure of pure NO$_2$ at 780 mbar. To avoid the dimer formation, the nozzle temperature is fixed at 80$^{\circ}{\rm C}$ where only $\sim$ 10 $\%$ species are N$_2$O$_4$. The latter has an extremely low excitation probability at 400 nm ($\leq 4\%$) \cite{N2O4exc}, so that it does not 
contribute to the detected signal. The harmonic signal is dispersed horizontally by a grazing incidence spherical grating with variable groove spacing and imaged by a detector consisting of microchannel plates, a phosphor screen and a charge-coupled device camera. 

\subsection{Experimental results}
As a first step we studied the picosecond dynamics of the diffracted and undiffracted signal. Similarly to ref. \cite{worner2011}, we observed picosecond transients reflecting the one-photon dynamics of NO$_2$ with a 1.8 $\pm$ 0.3 ps dissociation time. This indicates that the dominant process detected in our measurement is dictated by one-photon excitation. 

\begin{figure}
\begin{center}
\includegraphics*[width=0.5\textwidth]{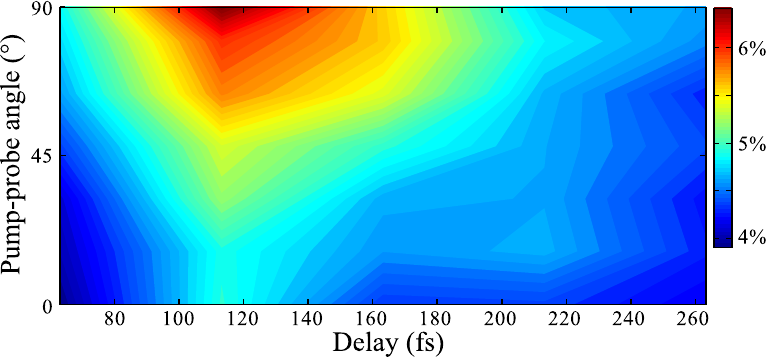}
\end{center}
\caption{First order diffraction efficiency for H15 as a function of the angle between pump and probe polarizations.}
\label{FigFemto_Angle}
\end{figure}

In order to study the electronic relaxation of photoexcited NO$_2$ molecules shown in Fig. \ref{FigPopAngles} and \ref{Fig_snapshot}, we have performed pump-probe measurements in the first few hundred femtoseconds following excitation. The pump pulses were 20 $\mu$J each and centered at 400 nm. First we determine angle between pump and probe polarizations that optimizes the contrast. Figure \ref{FigFemto_Angle} shows the first order diffraction efficiency $\eta$ for harmonic 15 as a function of delay and pump-probe polarization angle. The diffraction efficiency is defined as the fraction of the harmonic light contained within a single diffraction order :
 \begin{equation} 
 \eta=\frac{I_1}{I_{0}+I_{-1}+I_{1}} 
 \end{equation}
Monitoring this quantity rather than the raw diffracted signal enables lowering the effect of laser fluctuations and provides smoother signals. The diffraction efficiency shows a maximum around 120 fs. The contrast of this feature is maximized when the pump and probe polarizations are perpendicular to each other. All of the following discussion will therefore be with reference to this configuration. 

\begin{figure}
\begin{center}
\includegraphics*[width=0.5\textwidth]{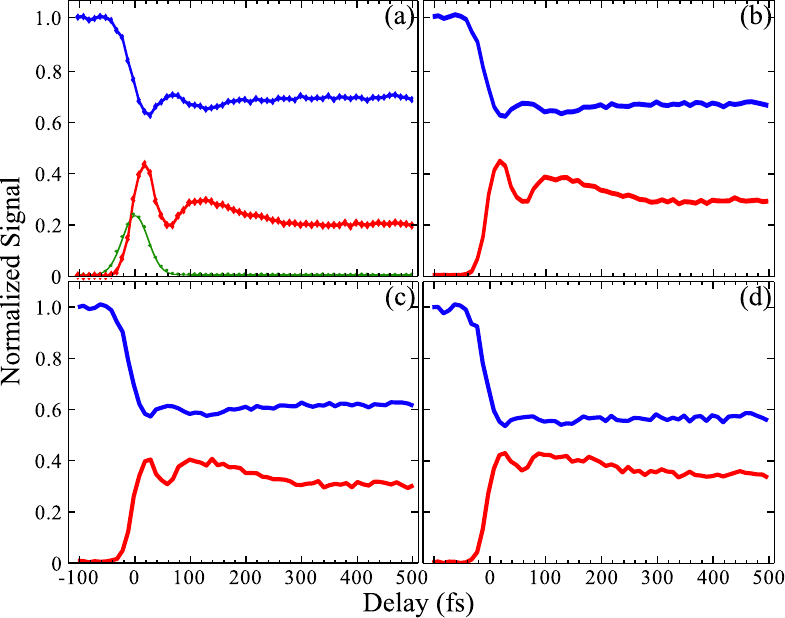}
\end{center}
\caption{Time dependency of the total harmonic signal (blue) and first order diffraction efficiency (red, mutiplied by 20) for harmonics 15 (a), 17 (b), 19 (c) and 21 (d). The pump pulses are 20 $\mu$J each, centered at 400 nm and polarized orthogonally to the probe. The cross correlation signal (panel (a): dots for the experimental points and green line for the gaussian fit) from harmonic 16 allows an accurate determination of the zero delay.}
\label{FigFemto}
\end{figure}

Figure \ref{FigFemto} shows the evolution of the total signal (blue) and first order diffraction efficiency (red) as a function of the pump-probe delay (10 fs steps), for harmonics 15 to 21. 
This result was obtained by adding up 5 consecutive scans, resulting in an accumulation over 6250 laser shots per delay. The zero pump-probe delay is determined from the even harmonics signal. These harmonics result from wave mixing between the pump (400 nm) and probe (800 nm) pulses, and thus provide a cross-correlation signal. The influence of the dynamics initiated by the pump pulse on even harmonics is taken into account to determine the zero delay. We assume that the even and odd harmonics are similarly affected by the change of the molecular geometry. The cross correlation signal is thus obtained by normalizing the signal of even harmonics by that of adjacent odds. The resulting signal has a symmetric shape, which can be fitted by a gaussian function to extract the zero delay. Note that this normalization shifts the zero delay by +10 fs. 

All harmonics experience a fast intensity decay within the first few tens of femtoseconds of interaction with the pump. After $t=20$ fs, the signal oscillates, presenting a maximum at 65 fs and a minimum at 130 fs. After that, it shows a slow increase, characteristic of the ongoing dissociation of a fraction of the excited molecules, since at 400 nm, with a 7 nm laser pump bandwidth, 25$\%$ of excited molecules lie above the dissociation limit. The diffraction efficiency shows an inverted behavior, with maxima at 15 fs and 120 fs, and a minimum at 58 fs. The contrast of diffraction efficiency is better than that of the total signal, in particular for highest harmonics. In parallel to the harmonic signal measurements, we measured the ionization yield from the generation jet (not shown here). The ion signal increases by a factor 1.3 between $t=-60$ fs and 60 fs and is stable afterwards.

Overall, these results are similar to the previously reported observation \cite{worner2011}, nonetheless they present a few differences. First, the diffraction efficiency measured here is 6 times lower than in the previous experiment. Second, dynamics appear as well in the undiffracted harmonic signal and not only in the diffracted light. Third, only two oscillations of the diffracted light are observed: a first sharp peak, and a second one with a steep leading edge and slow falling edge. Last but not least, the exact timing of the oscillations is different. In order to investigate the possible origins of these differences, we conducted a systematic study of the transient.

\begin{figure}
\begin{center}
\includegraphics[width=0.5\textwidth]{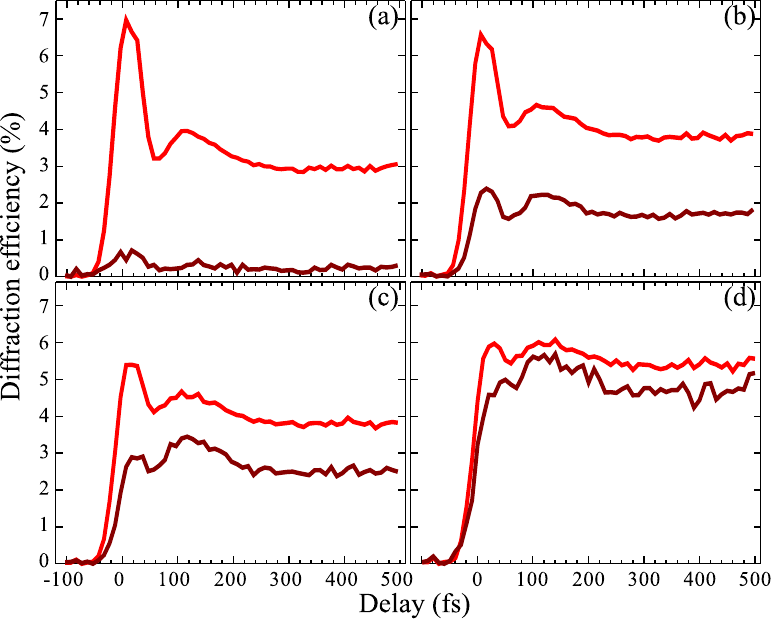}
\end{center}
\caption{Time dependency of the first order (light red) and second order (dark red, mutiplied by 10) diffraction efficiency for harmonics  15 (a), 17 (b), 19 (c) and 21 (d). The pump pulses are 35 $\mu$J each, are centered at 400 nm and are polarized orthogonally to the probe.}
\label{FigFemto_m1m2}
\end{figure}

When the pump energy is increased to 30 $\mu$J, second order diffraction peaks appear around the harmonics. Figure \ref{FigFemto_m1m2} compares the evolution of the first and second order diffraction efficiencies as a function of pump-probe delay. The dynamics observed for the second order diffraction is clearly the same as in the first order. This indicates that the second order diffraction peak originates from the saturation of the excitation rather than from a two-photon absorption process.  Indeed, with a $^2$B$_2 \leftarrow ^2$A$_1$ absorption cross-section \cite{Vandaele} $\sim$ 6$\times$10$^{-19}$ cm$^2$  and a pump waist size of $\sim 60$ $\mu$m, saturation is expected as soon as the pump energy reaches $\sim$25 $\mu$J per arm. 
For the first order, the results are very similar to those obtained at lower pump intensity in Fig. \ref{FigFemto}. While the relative heights of the peaks at 20 fs and 130 fs slightly differ between the two measurements, their positions are identical.  This shows that increasing the pump intensity does not affect the detected dynamics. We performed measurements varying the pump energy between 10 $\mu$J and 35 $\mu$J and did not observe significant changes in the temporal behavior of the first order diffracted signal. The first peak in the second order diffraction is slightly shifted to longer delays (Fig. \ref{FigFemto_m1m2}(d)), which could reflect the dynamics of the saturation process. 

\begin{figure}
\begin{center}
\includegraphics[width=0.5\textwidth]{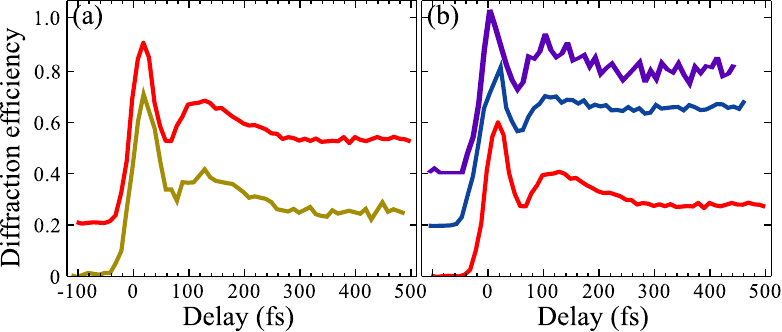}
\end{center}
\caption{First order diffraction efficiency in various excitation conditions. (a) Signal obtained using a warm (red) and cold (yellow) molecular beam. (b) Influence of the pump wavelength: 400 nm (red), 395 nm (blue) and 392 nm (purple). The signals are vertically shifted for sake of clarity.}
\label{FigTLambda}
\end{figure}

As the rotational temperature of the molecules may play a role in the measurement, in particular through the dynamics of the anisotropy induced by the pump excitation \cite{Blokhin,Raffael}, or the rovibrational and rovibronic couplings which are especially significant at long distances, we have compared the transients measured with two different gas sources. The first source is that used for all results presented so far: a 100 Hz General Valve pulsed source with a 100 $\mu$m nozzle diameter, backed by $\sim 780$ mbar of pure NO$_2$, and heated at 80$^\circ$C. In this configuration the laser is focused at a few hundred microns away from the nozzle. The second source is a 1 kHz Even-Lavie valve with a 250 $\mu$m conical nozzle, backed by a mixture of 16 bars of He and $\sim 800$ mbar of NO$_2$, and heated at 120$^\circ$C. In that configuration the laser is focused at $\sim 2$ mm away from the nozzle. We expect the rotational temperature to be significantly lower in the latter case. Figure \ref{FigTLambda}(a) compares the evolution of the diffraction efficiency for harmonic 17 using these two sources. While the ratio between the first and second maximum is different, 
which probably reflects slightly different laser pump conditions, the results are remarkably similar. 

Last, we studied the evolution of the transient as a function of the pump wavelength, by varying it between 392 and 400 nm (Fig. \ref{FigTLambda}(b)). The results are robust against such wavelength variations. 

In summary, none of the external parameters like pump intensity, wavelength and rotational temperature, can explain the differences observed between the data shown here and the ones published in ref. \cite{worner2011}. One possibility to explain the absence of oscillatory structure in the undiffracted HHG signal and more persistent oscillations seen in the earlier work could be the noise level, the statistics being improved in the present work due to the higher repetition rate of the laser system.

\section{Interpretation}
\begin{figure}
\begin{center}
\includegraphics[width=0.5\textwidth]{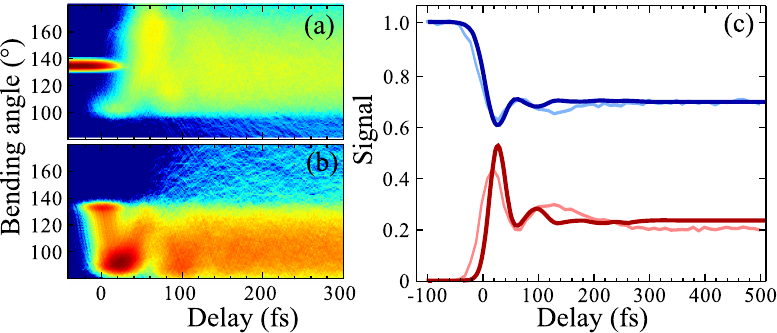}
\end{center}
\caption{Evolution of the calculated bending trajectory packet for the diabatic surfaces 1 (a) and 2 (b), in logarithmic scale. (c) Comparison between calculated (dark) and measured (light) high harmonic total signal (blue) and first order diffraction efficiency (red, multiplied by 20) for harmonic 15.}
\label{FigComparison}
\end{figure}

The experimental results show that the evolution of the diffraction efficiency presents robust characteristics: a maximum around 20 fs, a minimum around 60 fs, and a broader maximum around 120 fs. This behavior can be compared to that of the molecular dynamics described in Fig. \ref{FigPopAngles}. The minimum diffraction efficiency at 60 fs corresponds to a minimum in the population of the excited diabatic state, as well as to a maximum average bond angle of the molecule. In order to distinguish which of the two plays a dominant role, we performed simple calculations of the harmonic signal and diffraction efficiency. Since the tunnel ionization probabilities and recombination dipole moments have been calculated only as a function of the bending angle \cite{worner2011}, our model is one-dimensional: we only consider the influence of the motion along the bending coordinate by integrating the trajectory swarms over the stretching coordinates. The resulting evolutions are shown in Fig.\ref{FigComparison}(a-b).

\begin{figure}
\begin{center}
\includegraphics[width=0.5\textwidth]{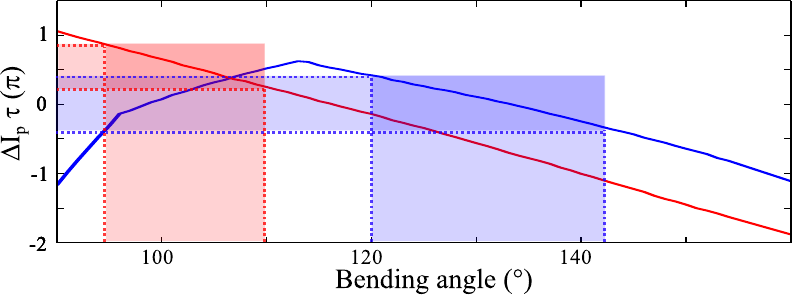}
\end{center}
\caption{Harmonic phase shift $\Delta I_p \tau$ in units of $\pi$, as a function of bending angle, for emission from PES 1 ($1^2A'$) (blue) and PES 2 ($2^2A' $)(red). The shaded areas correspond to the main part explored by the trajectories. }
\label{FigIpTau}
\end{figure}

In a three-step picture of HHG, the harmonic signal is determined by the strong field ionization rate of the molecule, the propagation of the electron in the continuum and the recombination dipole moment of this electron to the ion core. The ionization rates and complex recombination dipole moments were calculated in Ref \cite{worner2011} for the fundamental and first excited diabatic states, for three bond angles of the NO$_2$ molecules (85$^\circ$, 102$^\circ$ and 134$^\circ$), taking into account the different possible ionization channels. We use these results in our model and interpolate them to intermediate bond angles. The propagation of the electron in the continuum induces an additional phase shift in the harmonic emission from excited molecules with respect to ground state molecules, which corresponds to the phase $\varphi$ accumulated by the molecule between ionization and recombination : $\varphi=\Delta I_p\tau$ \cite{Kanai07,SmirnovaPNAS}, where $\Delta I_p$ is the difference between the ionization potential 
of the considered state and the unexcited ground state, and $\tau$ is the electron travel time, typically 1.2 fs for harmonic 17 for the experimental conditions. As the bond angle varies, the ionization potential changes because of the dramatic difference between the neutral and cationic potential surfaces. This can induce important variations of the harmonic phase. We calculate this phase by using the ionic potential curves from Ref \cite{arasaki10}. As shown in Fig. \ref{FigIpTau}, the phases vary linearly as a function of angle within the narrow range explored by the wavepacket, namely 120-142$^\circ$ for the ground electronic state and 90-110$^\circ$ for the excited state (shaded areas in Fig. \ref{FigIpTau}).

For each pump-probe delay $t$, the harmonic signal is calculated by coherently summing the contributions from the different bending angles $\theta$, for each molecular state. The undiffracted and first order diffracted signals are given by \cite{worner2011}:
\begin{eqnarray} 
\begin{split}
I_0(t)\propto \biggl|\int (1-r)d_g+r W_1(\theta,t) d_1(\theta)\\
+r W_2(\theta,t) d_2(\theta)d\theta \biggr|^2
\end{split} \\
\begin{split}
I_1(t)\propto \frac{1}{4} \biggl|\int r W_1(\theta,t) (d_1(\theta)-d_g)\\
+r W_2(\theta,t) (d_2(\theta) -d_g)d\theta\biggr|^2
\end{split}
\end{eqnarray}
where $r$ is the fraction of excited molecules, $d_g$ the harmonic dipole moment for the non excited molecules, $d_j(\theta)$ the complex dipole moment for emission from surface $j$ for the excited molecules, and $W_j(\theta,t)$ is the time-dependent bending trajectory packet on the $j$ surface as depicted in Fig. \ref{FigComparison} (a-b). Note that since the high harmonic signal is a coherent macroscopic emission, its amplitude is quadratically dependent on the population. The total signal and diffraction efficiency are $I_{tot}=I_0+2\times I_1$ and $\eta_1=I_1/I_{tot}$. Figure \ref{FigComparison}(c) shows the results obtained by assuming that $r=20\%$ of the molecules are excited, together with the experimental results. The results of the calculations have been convoluted by a gaussian function of 15 fs FWHM to take into account the duration of the harmonic generation process. The total signal is normalized to 1 at negative delays. The agreement is remarkably good, especially for the contrast, considering the simplicity of the model used for HHG calculations and the fact that we neglect the stretching motion. Both the fast dynamics and the average level of signals at long delays are well reproduced. Besides the slight temporal shift in the 
position of the first peak in the diffraction efficiency, the main difference between experiment and theory lies in the width of the second oscillation: the experimental results show a single broad secondary peak around 120 fs, associated to a dip in the total signal, while the calculated signal shows a peak at 100 fs followed by weaker replicas at 170 fs and 240 fs. This is probably an effect of the spreading of the packet along the stretching coordinates, which smoothes out the evolution at longer delays.

\begin{figure}
\begin{center}
\includegraphics[width=0.5\textwidth]{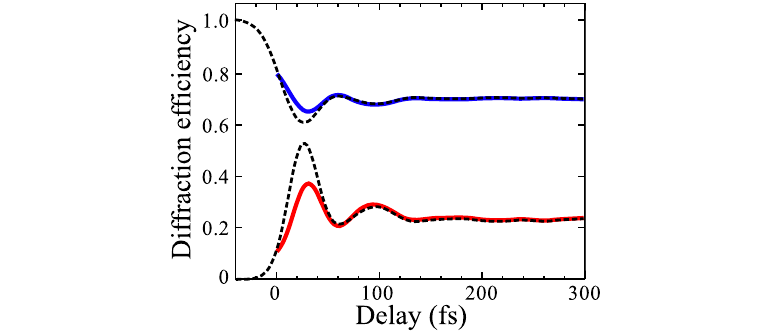}
\end{center}
\caption{Total harmonic signal (blue) and first order diffraction efficiency (red, multiplied by 20) obtained by imposing  constant equal populations for the two diabatic states. The black dotted curves are the results of the full calculation.}
\label{FigAnalysis}
\end{figure}
\begin{figure}
\begin{center}
\includegraphics[width=0.5\textwidth]{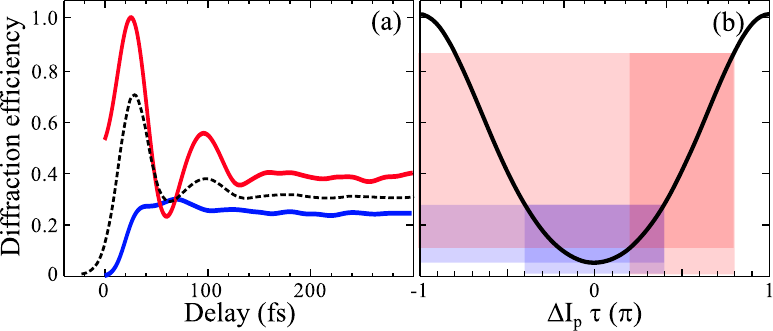}
\end{center}
\caption{(a) Partial contributions to the diffraction efficiency, considering only the trajectories on surface 1 ($P_1(t)=1$, $P_2(t)=0$, blue) or surface 2 ($P_1(t)=0$, $P_2(t)=1$,red). The black dashed line is the result of the full calculation. (b) Normalized diffraction efficiency as a function of the phase modulation $\Delta I_p \tau$ in units of $\pi$. The shaded areas correspond to the main part explored by the trajectories.}
\label{FigAnalyzeBending}
\end{figure}

In order to ascertain the origin of the transients observed on diffracted signals,
we cancel out the effect of the population transfers by normalizing the packets at each delay: $P_1(t)=\int W_1(\theta,t)d\theta=1/2$ and $P_2(t)=\int W_2(\theta,t) d\theta=1/2$. This enables us to isolate the influence of the bending motion.  The results are remarkably similar to those obtained with the full calculation (Fig. \ref{FigAnalysis}). 
The influence of the packet dynamics can be further disentangled by calculating the contribution of each packet alone (Fig. \ref{FigAnalyzeBending} (a)), setting $P_1(t)=1$ and $P_2(t)=0$ (blue) or $P_1(t)=0$ and $P_2(t)=1$ (red). This reveals that the main origin of the observed experimental modulations is the bending motion of the trajectory packet on the excited diabatic surface 2. This is surprizing, given the fact that the extension of the bending motion is more important on the diabatic surface 1 than on 2. As the packets move, the harmonic emission is mainly modulated by the phase accumulated by the molecules during the electron travel in the continuum, $\Delta I_p \tau$ (Fig. \ref{FigIpTau}). This phase, which appears in the complex dipole moments, typically varies between $-0.4\pi$ and $0.4\pi$ during the packet motion on surface 1, and between $0.2\pi$ and $0.8\pi$ during the packet motion on surface 2. Figure \ref{FigAnalyzeBending}(b) shows the evolution of the diffraction efficiency as a function of $\Delta I_p \tau$. It is clear from this graph that the packet motion on surface 2 will induce a much stronger diffraction signal as well as a deeper modulation. This shows that similar variations of $\Delta I_p \tau$ as a function of the reaction coordinates can have very different impact on the signal depending on the absolute value of the phase. For instance an additional recombination phase in the harmonic emission could strongly modify the situation and make the dynamics on surface 1 dominate the signal. This issue is for instance illustrated in the different contrast observed as a function of the harmonic order for the first and second order diffraction efficiency. The sensitivity of high harmonic spectroscopy to specific dynamics will thus depend on the considered system and must be carefully studied case by case.

\begin{figure}
\begin{center}
\includegraphics[width=0.5\textwidth]{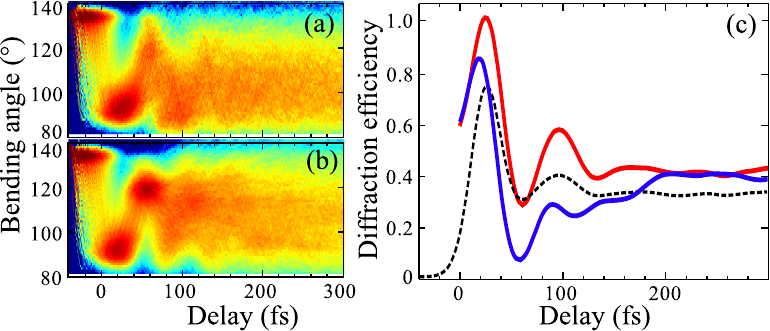}
\end{center}
\caption{Temporal evolution of the calculated trajectory packet on the excited diabatic surface 2 with (a) and without (b) the coupling to the ground state. (c) First order diffraction efficiency induced by the trajectories on surface 2, with (red) and without (blue) the coupling to the ground state. The black dashed line is the result of the full calculation. }
\label{FigCoupling}
\end{figure}

Since the signal is dominated by the vibrational motion on the excited diabatic surface, one may wonder what is the influence of the conical intersection on this signal. To estimate this, we calculated the molecular dynamics on surface 2 in the absence of electronic coupling to surface 1. The results are shown in Fig. \ref{FigCoupling}(b) and compared to the results including the coupling but keeping the population constant $P_2(t)=1$ (Fig.\ref{FigCoupling}(a)). The results are similar until 20 fs delay. This means that the first passage through the conical intersection, around 10 fs delay, has only little effect on the  structure of the trajectory packet. By contrast at 60 fs there is a strong transfer to the ground diabatic state (Fig \ref{Fig_snapshot}) which alters the packet, depleting the highest bending angles. This is followed by a transfer back to the excited state around 100 fs. Figure \ref{FigCoupling}(c) compares the high harmonic diffraction efficiencies obtained with (red) and without (blue) the coupling. The results are quite different. In particular they indicate that the second oscillation observed around 100 fs in the experimental signal is due to the transfer through the conical intersection from the ground to the first excited diabatic state. While the change in the overall population of the excited state has only little effect on the signal (Fig. \ref{FigAnalysis}), the packet structure and dynamical changes induced by transfers through the conical intersection are clearly reflected in the signal. 

The present results have to be compared to recent calculations based on wavepacket propagation \cite{Kraus12PRA}. In a first 1D calculation, the bending wavepackets were modelled as Gaussian functions of constant width, with their centers following the angular position of the full 3D wavepackets center. The harmonic emission was calculated using the same procedure as described here, i.e. taking into account the phase modulations induced by the vertical ionization potential changes, and the changes in the harmonic dipole moments due to the change in molecular geometry. This calculation resulted in strongly contrasted modulations of the diffracted signal as a function of delay. In a second calculation, the wavepacket motion was neglected and only the influence of the populations was retained. The resulting modulation of the diffracted light was weaker than in the 1D case. Last, a 3D calculation taking into account the evolution of the molecular wavepackets along the three dimensions was performed, assuming constant harmonic dipole moments and phases except for from the $I_p\tau$ phase. The results hardly show any modulation in the harmonic signal \cite{Kraus12CP}. Since the model neglecting wavepacket motion and taking into account only the population evolutions gave the best qualitative agreement with their experiments, Kraus \textit{et al.} concluded that the experimental signal reflected the variation of the diabatic state populations, the phase modulations $I_p\tau$ associated to the vibrational motion being quickly washed out by wavepacket spreading in the three dimensions. This conclusion is quite different from that drawn from our 1D model, in which we find that the population transfer only has a weak effect on the detected signal. It is interesting to note that for the 1D models, the proper treatment of the extension of the wavepacket along the bending angle (as done in equation 9 and 10 in the present paper) without the assumption of a Gaussian shape as done in \cite{Kraus12CP} leads to a better agreement with experiments. 

If only the populations or vertical ionization potential were probed by high harmonic spectroscopy, the measurement should reveal the same dynamics irrespective to the pump-probe angle. This is not the case: the dynamics is much more contrasted for orthogonal pump and probes, as shown in Fig. \ref{FigFemto_Angle}. This is a consequence of the angular dependence of the harmonic signal with respect to the z-molecular axis: the harmonic dipole moments can indeed play a major role in the outcome of the measurement. The fact that the 1D model used here reproduces better the experiment than the 3D model previously used in \cite{Kraus12PRA} assuming constant harmonic dipole moments for all bending angles and bond lengths suggests that the harmonic dipole moments present significant variations along the stretching coordinates, such that the complete 3D averaging is closer to the complete 1D calculation than to the partial 3D calculation. A complete 3D calculation, including the evolution of the harmonic dipole moments along all molecular coordinates, is required to confirm this statement. 

\section{Conclusion}
High harmonic transient grating spectroscopy experiments in photoexcited NO$_2$ molecules show modulations of both the total harmonic signal and first order diffraction efficiency within the first 200 fs following the pump pulse. These can be remarkably well reproduced by combining Trajectory Surface Hopping calculations of the reaction dynamics around the conical intersection and a simple model of high harmonic generation. The analysis reveals that the dominant information encoded in the transient grating signal is the molecular vibration on the first excited diabatic state. This sensitivity results from the modulation of the vertical ionization potential as the molecule vibrates, which induces a phase modulation of the harmonic emission. The vibrational motion in the ground diabatic state also induces a significant phase modulation. However this modulation is centered on zero (the zero phase shift corresponding to the equilibrium geometry of the molecule in the ground state), such that it has only little effect on the measured signal. 

These results show the importance of the harmonic phase in the nature of the information encoded in high harmonic transient grating spectroscopy experiments. Indeed, an additional phase shift in the harmonic emission could invert the situation and make the signal most sensitive to the vibration in the ground diabatic state. Since the phase of the recombination dipole moments in HHG can show significant structures as a function of  harmonic order \cite{Smirnova09b,Le09}, it would be interesting to extend this study to longer driving wavelengths which produce much broader harmonic spectra. The transient grating signal could then encode different dynamics (ground diabatic state vibrations, first excited diabatic state vibrations, or population transfer through the conical intersection) depending on the harmonic order, giving a full picture of the ongoing dynamics. 

We acknowledge R. Schinke for providing the PESs used in the TSH calculations. We thank R. Bouillaud and C. Medina for their technical assistance. We acknowledge financial support by the ANR (ANR-08-JCJC-0029 HarMoDyn), the Conseil Regional Aquitaine (20091304003 ATTOMOL and 2.1.3-09010502 COLA2 projects), and the European Union through the MOLCOTUV and ITN-ICONIC contracts.

\bibliographystyle{aip}
\end{document}